\documentclass[aps,prc,a4paper,nofootinbib,
preprintnumbers,superscriptaddress,twocolumn,showpacs,showkeys]{revtex4}

\usepackage{amsmath}
\usepackage{amssymb}
\usepackage{bm}
\usepackage{graphicx}
\usepackage{color}

\definecolor{babypink}{rgb}{0.96, 0.76, 0.76}
\definecolor{darkpastelgreen}{rgb}{0.01, 0.75, 0.24}

\begin{document}

\title{Photons from the Early Stages of Relativistic Heavy Ion Collisions}

\author{L. Oliva}
\affiliation{Department of Physics and Astronomy, University of Catania, Via S. Sofia 64, I-95123 Catania}
\affiliation{INFN-Laboratori Nazionali del Sud, Via S. Sofia 62, I-95123 Catania, Italy}

\author{M. Ruggieri}%\email{marco.ruggieri@ucas.ac.cn}
\affiliation{College of Physics, University of Chinese Academy of Sciences, Yuquanlu 19A, Beijing 100049, China}

\author{S. Plumari}
\affiliation{Department of Physics and Astronomy, University of Catania, Via S. Sofia 64, I-95123 Catania}
\affiliation{INFN-Laboratori Nazionali del Sud, Via S. Sofia 62, I-95123 Catania, Italy}

\author{F. Scardina}
\affiliation{Department of Physics and Astronomy, University of Catania, Via S. Sofia 64, I-95123 Catania}
\affiliation{INFN-Laboratori Nazionali del Sud, Via S. Sofia 62, I-95123 Catania, Italy}

\author{G. X. Peng}
\affiliation{College of Physics, University of Chinese Academy of Sciences, Yuquanlu 19A, Beijing 100049, China}
\affiliation{Theoretical Physics Center for Science Facilities, Institute of High Energy Physics, Beijing 100049, China}

\author{V. Greco}
\affiliation{Department of Physics and Astronomy, University of Catania, Via S. Sofia 64, I-95123 Catania}
\affiliation{INFN-Laboratori Nazionali del Sud, Via S. Sofia 62, I-95123 Catania, Italy}
\affiliation{College of Physics, University of Chinese Academy of Sciences, Yuquanlu 19A, Beijing 100049, China}

\begin{abstract}
We present results about photons production in relativistic heavy ion collisions. 
The main novelty of our study is the calculation of the contribution of the early stage photons
to the photon spectrum. 
The initial stage is modeled by an ensemble of classical gluon fields which decay to a quark-gluon plasma
via the Schwinger mechanism, and the evolution of the system is studied by coupling
classical field equations to relativistic kinetic theory; photons production is then computed by 
including the pertinent 
collision processes into the collision integral. We find that the contribution of the early stage photons 
to the direct photon spectrum is substantial for $p_T \approx 2$ GeV and higher,
the exact value depending on the collision energy;
therefore we identify this part of the photon spectrum as the sign of the early stage. 
Moreover, the amount of photons produced during the early stage is not negligible with respect to
those produced by a thermalized quark-gluon plasma: we support the idea that there is no dark age
in relativistic heavy ion collisions.
\end{abstract}

%\vspace{10pt}
%\begin{indented}
%\item[]February 2016
%\end{indented}

\pacs{25.75.-q, 25.75.Ld, 25.75.Nq, 12.38.Mh}
\keywords{Relativistic heavy ion collisions, Quark-gluon plasma, 
Relativistic Transport Theory, Photon emission.}

\maketitle

\section{Introduction}
Photons are important probes of the system produced in
relativistic heavy ion collisions (RHICs), offering an useful way to investigate the
pre-equilibrium stage, the quark-gluon plasma (QGP) and the hadronic phase.
As a matter of fact, photons are emitted during the whole lifetime of the system produced by the collisions, and 
because their mean free path is much larger than the collision volume,
they leave the system undisturbed. For this reason it is often said that they bring to the detectors the
informations about the particular state that has produced them.
Photon production in RHICs has been studied extensively within recent years,
see \cite{Shen:2013vja,Liu:2007tw,Liu:2008eh,Liu:2014fya,paquet2016,arnold2001,
Turbide:2003si,Linnyk:2015tha,Linnyk:2013wma,Linnyk:2015rco,Berges:2017eom,
Kapusta:1991qp,Baier:1991em,Baier:1991bz,
Ayala:2016lvs,Vovchenko:2016mtf,Chiu:2012ij,Ghiglieri:2016tvj,Ghiglieri:2013gia,
Greif:2016jeb,Iatrakis:2016ugz,Chatterjee:2013naa,Chatterjee:2012dn,Liu:2015vst,Liu:2016ijc,Liu:2012ax}
and references therein. 
 
In the lifetime of the fireball produced in RHICs it is possible 
to distinguish among direct photons, namely those arising from collision processes,
and decay photons that are instead produced by hadron decays. 
Direct photons are then mainly split into prompt photons, produced by primordial 
scatterings among the nucleons, and thermal photons, that instead are produced by a thermalized
QGP and hadron gas. Both thermal and prompt photons have been already intensively studied. 
%{\color{blue}
It should be remarked however that the problem of photon production from a thermalized
quark-gluon plasma is still not solved completely: as a matter of fact, the most used rate for a thermalized
quark-gluon plasma \cite{arnold2001} corresponds to a weak coupling limit result
(next-to-leading order corrections to this result exist~\cite{Ghiglieri:2013gia} but lead to at most
a $20\%$ shift upwards of the rate);
the question of the thermal emission of photons from a strongly coupled plasma,
like the one produced by RHICs, has only recently been 
addressed by means of holographic techniques~\cite{Iatrakis:2016ugz}.
Therefore it is fair to say that the
problem is not yet completely understood. %}

Besides, the study of the photon emission in the 
early non-equilibrium stage of HICs is still not complete, since only a very small
amount of work has been devoted to this subject \cite{Berges:2017eom,Linnyk:2015tha,Linnyk:2013wma,Linnyk:2015rco,
Vovchenko:2016mtf,Chiu:2012ij,Greif:2016jeb}.
We aim to fill this gap here, by presenting results about photon production 
considering pre-equilibrium photons on the same footing of thermal photons.  

Our main theoretical scheme consists of relativistic transport theory coupled
to the dynamics of a classical color field that corresponds to the initial gluon-dominated stage, and eventually
evolves to QGP.
This theoretical model has been used to study the evolution of the early stage of RHICs
\cite{Ryblewski:2013eja,Ruggieri:2015yea} giving a picture that agrees qualitatively,
and to some extent also quantitatively, with the one obtained by means of classical Yang-Mills calculations
for what concerns the evolution of the pressures in the system \cite{Gelis:2013rba}.

Although the initial state is represented by a classical
field that mimics the Glasma \cite{Lappi:2006fp}, the decay of the field
produces quickly quarks and gluons that scatter and create photons even when the system
is not a thermalized QGP. We implement photons production by means of a collision integral,
therefore we do not follow the common strategy used in previous calculations
in which one has to assume local thermalization 
and integrate the rates over the spacetime volume of the fireball.
This is the advantage of using relativistic transport theory,
which allows us to study photons production also in the early stages 
where hydrodynamics cannot be properly (or judiciously) applied.
Nonetheless, we will see that in thermal QGP phase the two approaches gives nearly identical results.

With respect to previous calculations based on transport theory
\cite{Linnyk:2015tha,Linnyk:2013wma,Linnyk:2015rco}, the main novelty that we bring by our study 
is the clear identification of the contribution of the early stage of RHICs to the direct photon spectrum,
looking for both the relative abundance of the photons produced and the momentum region of the photon 
spectrum that takes the main contribution from the early stage. We will find that the
early stage is quite efficient in producing photons, therefore our results support the absence 
of a dark age in RHICs.
Photons from a thermalizing early stage have been also studied very recently in
\cite{Berges:2017eom}, where the bottom-up thermalization scenario of \cite{Baier:2000sb}
has been adopted;  classical-statistical simulations have shown that 
bottom-up is the right thermalization scenario \cite{Berges:2013eia,Berges:2013fga}
and it extrapolates to finite couplings quite well \cite{Kurkela:2015qoa,Keegan:2016cpi}. 
Our results agree with the importance of 
the early stage photons production in RHICs already highlighted in \cite{Berges:2017eom};
with respect to \cite{Berges:2017eom}, our main novelty is to implement photon production
by a code based on relativistic transport theory and set up to follow 
the dynamical evolution of the system produced in RHICs, from the early stage
up to the freezout, 
thus allowing a more direct  link to the observables of RHICs. 
In the bottom-up scenario, Bose-Einstein enhancement factors
in the collision integral are potentially important due to large gluon occupation numbers in the initial stage;
these have been considered in \cite{Berges:2017eom}, while for simplicity
we have not included them in our calculations although within transport theory they can
be implemented \cite{Scardina:2014gxa}, leaving their inclusion to future works. 

%{\color{blue}
Photons from the pre-equilibrium stage have been studied very recently  within
another relativistic transport code~\cite{Greif:2016jeb}, where a gluon dominated initial state is considered 
and quarks are produced by means of inelastic scatterings. The main difference between our work and \cite{Greif:2016jeb}
is that in the former, quarks and gluons are produced on the same footing by the decay of the initial classical
gluon field, that results in a quicker photon production in the early stage.
The gluon dominated initial stage supported in \cite{Greif:2016jeb} agrees with the one
advertised in \cite{Vovchenko:2016mtf,Liu:2012ax} where it has been also suggested that the delayed quark emission
can help to explain the large elliptic flow of photons, in agreement with
the analysis of~\cite{Liu:2014fya}. Whether the delayed photon emission can really help to understand
the photon $v_2$ puzzle is still an open problem, see \cite{Linnyk:2015rco} for a review. 
The problem of the direct photon elliptic flow has been also addressed with other relativistic transport calculations~\cite{Linnyk:2015tha}.
We will not consider %in detail 
the $v_2$ of photons in this article because our goal is to
discuss the photon production in the pre-equilibrium stage, a problem that is equally
important and nowadays still under debate.
%}

The plan of the article is as follows. In Section II we review the abelian flux tube model
that we use for the initial condition, as well as our implementation of transport theory
and the photon rate that we implement in the collision integral.
In Section III we summarize our main results on the photon spectrum and photon abundancy.
Finally in Section IV we draw our conclusions.

\section{Initial condition, its evolution and photon production}

\subsection{Abelian Flux Tube model}
In this subection, we summarize the Abelian Flux Tube model (AFTm),
that we use to define an initial condition in our simulations based on classical gluon fields, 
as well as the base for the evolution of this field configuration to QGP, see  
\cite{Casher:1978wy,Glendenning:1983qq,Bialas:1984wv,Bialas:1984ap,Bialas:1985is,
Gyulassy:1986jq,Gatoff:1987uf,Elze:1986qd,Elze:1986hq,Bialas:1986mt,
Florkowski:2003mm,Bajan:2001fs,Bialas:1989hc,Dyrek:1988eb,
Bialas:1987en,Ryblewski:2013eja,Florkowski:2014ska,Florkowski:2010zz,
Ruggieri:2015yea,Ruggieri:2016ckn} for details.

The main idea of the AFTm is to replace the Glasma with a simpler initial classical color field configuration,
in which one considers, in its simplest realization, only the electric part of the color field, which decays
into QGP by means of the Schwinger mechanism; the pair creation that occurs locally changes the local dipole
moment of the system, creating a displacement current whose backreaction on the evolution of the field
is taken into account properly (see below). 
Besides, it is assumed that the classical field equations are abelian, namely the covariant derivatives
in the QCD field equations are replaced by ordinary  derivatives. 
We would like to observe, however, that even if the model is named abelian, such nomenclature
simply refers to the fact that in the evolution equation for the classical field, self-interaction terms
coming from non vanishing structure constants of the color gauge group are neglected \cite{Florkowski:2010zz}. 
However interactions among the classical field and gluons are still present in this calculations,
thanks both to the Schwinger effect which produces charged gluons, and to conduction currents which
affect the evolution of the field, see the next Section for more details. 
%{\color{blue}
It would be certainly interesting to include the effects of a magnetic color field
as well as of the nonabelian terms in the classical equations of motion%:
; the nonabelian effects have been investigated in \cite{Voronyuk:2015ita} in the SU(2) case.
The upgrade of our simulation code to this more realistic initial condition and
early stage dynamics is in progress and the results will be the subject of
forthcoming publications. 
%}

In this work, we assume that the initial color electric field is
longitudinal, while transverse components of the field develop due to transverse color currents.
Assuming massless quanta, the number of  pairs per unit of spacetime
and invariant momentum space produced by the decay of the electric field by the Schwinger effect 
is~\cite{Ryblewski:2013eja}
\begin{equation}
\frac{dN_{jc}}{d\Gamma}\equiv
p_0\frac{dN_{jc}}{d^4x d^2 p_T dp_z} = {\cal R}_{jc}(p_T)
\delta(p_z)p_0,
\label{eq:SF}
\end{equation}
with
\begin{equation}
{\cal R}_{jc}(p_T) =\frac{{\cal E}_{jc}}{4\pi^3}
\left|
\ln
\left(
1\pm e^{-\pi p_T^2/{\cal E}_{jc}}
\right)
\right|,
\label{eq:RpT}
\end{equation}
the plus (minus) sign corresponding to the creation of a boson (fermion-antifermion) pair.
In this equation $p_T$, $p_z$  refer to
each of the two particles created by the tunneling process;
${\cal E}_{jc}$ is the effective force which acts on the tunneling pair and it depends
on color and flavor; it can be written as
\begin{equation}
{\cal E}_{jc} = \left(
g|Q_{jc}E| - \sigma_{j}
\right)
\theta
\left(
g|Q_{jc}E| - \sigma_{j}
\right),
\label{eq:sigma}
\end{equation}
where $E$ stands for the magnitude of the color field,
$\sigma_j$ denotes the string tension depending on the kind of flavor considered.
Moreover $p_0=\sqrt{p_T^2 + p_z^2}$ corresponds to the single particle kinetic energy.

The $Q_{jc}$ are color-flavor charges which, in the case of quarks,
correspond to the eigenvalues of the $T_3$ operator:
\begin{equation}
Q_{j1}=\frac{1}{2}~,~~~~Q_{j2}=-\frac{1}{2}~,~~~~Q_{j3}=0~,~~~~~j=1,N_f;
\end{equation} 
for antiquarks, corresponding to negative values of $j$, the color-flavor charges
are just minus the corresponding charges for quarks. 
Finally for gluons (which 
in our notation correspond to $j=0$) the charges are obtained
by building gluons up as the octet of the $3\otimes \bar 3$
in color space; in particular \cite{Florkowski:2010zz,Ruggieri:2015yea}
\begin{equation}
Q_{01}=1~,~~~~Q_{02}=\frac{1}{2}~,~~~~Q_{03}=-\frac{1}{2},
\end{equation} 
and $Q_{04}=-Q_{01}$, $Q_{05}=-Q_{02}$, $Q_{06}=-Q_{03}$.

\subsection{Relativistic Transport Theory}

Our calculation scheme is based on the Relativistic Transport Boltzmann equation, namely
\begin{equation}
\label{Boltzmann}
\left( p^{\mu}\partial_{\mu} + 
g Q_{jc} F^{\mu\nu}p_{\nu}\partial_{\mu}^{p} \right)f_{jc}(x,p)=\frac{dN_{jc}}{d\Gamma}+{\cal C}_{jc}[f],
\end{equation}
where $f_{jc}(x,p)$ is the distribution function for flavour $j$ and color $c$, $F^{\mu\nu}$ is the field strength tensor. 
On the right hand side we have the source term $dN/d\Gamma$ which describes the creation of quarks, antiquarks and gluons due to the decay of the color electric field and ${\cal C}[f]$ which represents the collision integral.
Considering only $2\to 2$ body elastic scatterings, the collision integral can be written as:
\begin{eqnarray}
{\cal C}[f]&=& 
 \int\frac{1}{2E_1 }
 \frac{d^3p_2}{2E_2(2\pi)^3}  \frac{d^3p_{1^\prime}}{2E_{1^\prime}(2\pi)^3}  
 \frac{d^3p_2^\prime}{2E_2^\prime(2\pi)^3}\nonumber\\
&&\times(f_{1^\prime}f_{2^\prime} - f_1 f_2) |{\cal M}|^2\delta^4(p_1 + p_2 - p_{1^\prime} - p_{2^\prime}),\nonumber\\
&&
\end{eqnarray}
where we omit flavour and color indices for simplicity,  
${\cal M}$ is the transition matrix for the elastic process linked to the differential cross section 
through $|{\cal M}|^2=16\pi s^2d\sigma/dt$, being $s$ the Mandelstam variable.
In our simulations we solve numerically Eq.~\eqref{Boltzmann} 
using the test particles method and the collision integral is computed 
using Monte Carlo methods based on the stochastic interpretation 
of transition amplitude \cite{Xu:2008av,Xu:2007jv,Bratkovskaya:2011wp,Ferini:2008he,
Plumari_BARI,Plumari:2012xz,Ruggieri:2013bda,Ruggieri:2013ova,Plumari:2012ep}.

The evolution of the electric color field is given by
\begin{equation}
\bm\nabla\cdot\bm E  = \rho,~~~
\frac{\partial \bm E}{\partial t} = -\bm j,
\label{eq:MAX_tt}
\end{equation}
where $\rho$ corresponds to the color charge density,
\begin{equation}
\rho = g\sum_{j,c}Q_{jc}\int d^3\bm p  f_{jc}(p),
\end{equation}
with $j,c$ standing for flavor and color respectively; the sum in the above equation runs over quarks, antiquarks and gluons.
On the other hand, $\bm j$ corresponds to the color electric current
which is given by the sum of two contributions, that is
\begin{equation}
j = j_M + j_D.
\end{equation}
Here $j_M$ is a colored generalization of the usual electric current density
which in a continuum notation is given by
\begin{equation}
j_M^\mu = g\sum_{j,c}Q_{jc}\int\frac{d^3\bm p}{p_0}p^\mu f_{jc}(p)~.
\end{equation}
The term $j_D$ is called the displacement current, arising from the 
change in time of the dipole moment of the medium induced
by pairs pop-up via the Schwinger mechanism,
in the same way a time variation of the local dipole moment in a medium gives rise
to a change in the local electric field~\cite{MAxwell}.
We can write~\cite{Ryblewski:2013eja}
\begin{equation}
j_D =\sum_{j=0}^{N_f}\sum_{c=1}^3
\int\frac{d^3\bm p}{p_0}
 \frac{dN_{jc}}{d\Gamma} 
\frac{2p_T}{E},
\label{eq:jD}
\end{equation}
where $N_f$ corresponds to the number of flavors in the calculation.
The color charge and current densities depend on the particle distribution
function: hence they link the field equations~\eqref{eq:MAX_tt} to the kinetic 
equation~\eqref{Boltzmann}. 
We solve self-consistently the field and kinetic equations,
taking into account the back reaction of particle production and propagation
on the color field.

At variance with the standard use of transport theory, in which one fixes a set of microscopic processes into the collision integral, we have developed an approach that
fixes the total cross section in order to have the wanted $\eta/s$ of the system. 
By means of this scheme we are able to use the Boltzmann equation to simulate the dynamical evolution of a fluid with specified shear viscosity, in analogy to what is done within hydrodynamical 
simulations \cite{Denicol:2014tha,Huovinen:2008te,El:2009vj}.

We use the Chapman-Enskog \cite{CE} approach to relate shear viscosity to temperature, cross section and density
which is in agreement with Green-Kubo correlator results \cite{Plumari:2012ep}.
Therefore, we fix $\eta/s$ and compute the pertinent total cross section by mean of the relation
\begin{equation}
\sigma_{tot}=\frac{1}{5}\frac{T}{\rho \, g(a)} \frac{1}{\eta/s}, 
\label{eq:sigmaTT}
\end{equation}
which is valid for a generic differential cross section $d\sigma/dt \sim \alpha_s^2/(t-m_D^2)^2$
as proved in~\cite{Plumari:2012ep}.
In the above equation $a=m_D/2T$, with $m_D$ the screening mass regulating the angular dependence
of the cross section, while  
\begin{eqnarray}
g(a)&=&\frac{1}{50}\! \int\!\! dyy^6
\left[ (y^2{+}\frac{1}{3})K_3(2 y){-}yK_2(2y)\right]\!
h\left(\frac{a^2}{y^2}\right),\nonumber\\
&&
\label{g_CE}
\end{eqnarray}
where $K_n$ is the Bessel function and the function $h$ relates the transport cross section to the total one $\sigma_{tr}(s)= \sigma_{tot} \, h(m_{D}^2/s)$
being  $h(\zeta)=4 \zeta ( 1 + \zeta ) \big[ (2 \zeta + 1) ln(1 + 1/\zeta) - 2 \big ]$. 
The $g(a)$ is the proper function accounting for
the correct relaxation time $\tau_{\eta}^{-1}=g(a) \sigma_{tot} \rho$ associated
to the shear viscosity transport coefficient. 
For isotropic cross section, i. e. $m_D\to \infty$, the function $g(a)$ is equal to $2/3$
%The maximum value of $g(a)$, namely $g( m_D \rightarrow \infty)=2/3$, 
%is reached for isotropic cross section 
and Eq.(\ref{eq:sigmaTT}) reduces to the relaxation time approximation
with $\tau^{-1}_{\eta} =\tau^{-1}_{tr} = \sigma_{tr} \rho$, while for finite value of $m_D$, which means anisotropic scatterings, $g(a)<2/3$.
We notice that, in the regime where viscous hydrodynamics applies, the specific microscopic details of the cross section are irrelevant, and our approach is an effective way to employ transport theory to simulate a fluid at 
a given $\eta/s$.

\subsection{Photon production rate}
\begin{figure*}[t!]
\begin{center}
\includegraphics[scale=0.55]{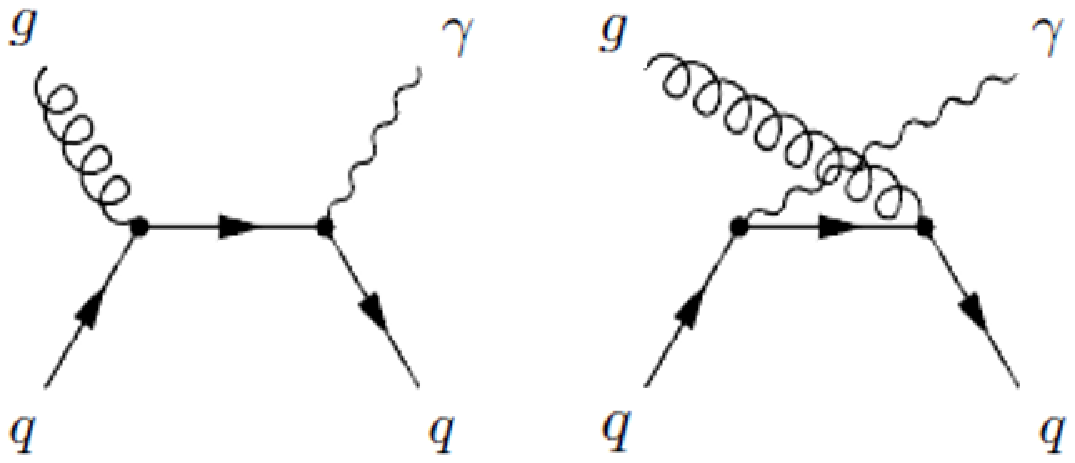}~~~\includegraphics[scale=0.55]{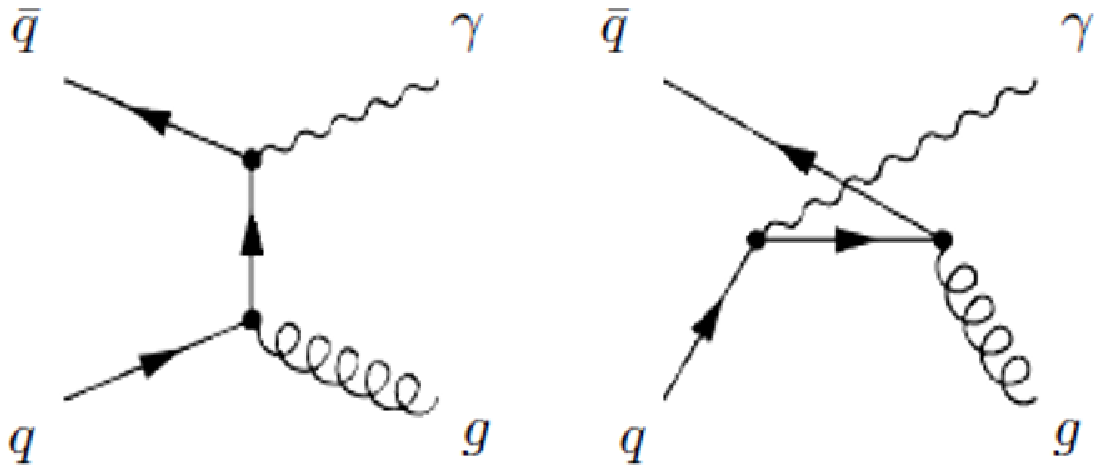}\\
{\bf(a)}~~~~~~~~~~~~~~~~~~~~~~~~~~~~~~~~~~~~~~~~~~~~~~~~~~~~~~~~~~~{\bf(b)}
\caption{\label{Fig:a1}Microscopic processes implemented in the collision integral:  
{\bf(a)} stands for Compton scattering and {\bf(b)} corresponds to pair annihilation.}
\end{center}
\end{figure*}

\begin{figure}[t!]
\begin{center}
\includegraphics[scale=0.35]{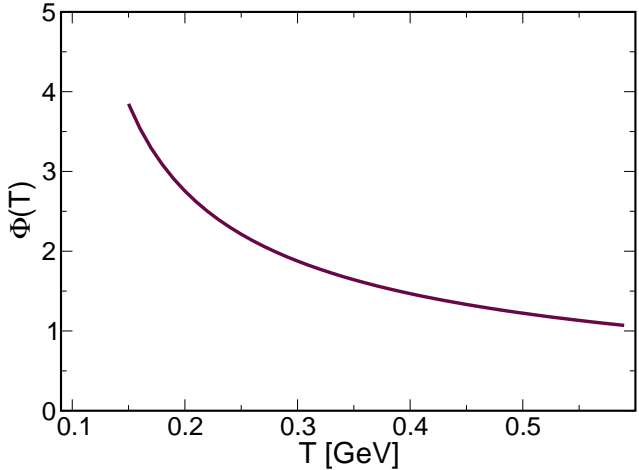}
\caption{\label{Fig:phi}Function $\Phi(T)$ appearing in Eqs.~\eqref{eq:1a} and~\eqref{eq:2a},
versus the effective temperature $T=E/3N$.}
\end{center}
\end{figure}

In this study, we implement photon production by adding the  $2\rightarrow2$ standard
processes of Compton scattering and quark-antiquark annihilation in the collision integral, 
see Fig.~\ref{Fig:a1}. 
The differential cross sections for the processes are given by
\begin{eqnarray}
\frac{d\sigma^{Compton}}{dt}&=&-\frac{\pi\alpha\alpha_s}{3s^2}\frac{u^2+s^2}{us},
\label{eq:1}
\\
\frac{d\sigma^{annihil}}{dt}&=&\frac{8\pi\alpha\alpha_s}{9s^2}\frac{u^2+t^2}{ut},\label{eq:2}
\end{eqnarray}
where $s,t,u$ represent the standart Mandelstam variables. 
In these equations $\alpha_s$ corresponds to the strong coupling, that we take running according to the
one-loop QCD $\beta-$function, the sliding scale being the local temperature of the fluid. 
The photon production rate
that would result by considering only the processes in Fig.~\ref{Fig:a1} has been computed
in \cite{Kapusta:1991qp,Baier:1991em,Baier:1991bz}; however, it is well known that the infrared enhancement of $2\rightarrow3$ processes is important
and makes these processes as important as the $2\rightarrow2$ ones, although 
these processes would appear to be suppressed by a naive coupling power counting \cite{arnold2001}.
The  $2\rightarrow3$ processes lead to an increase of the scattering rate with respect to the one 
obtained considering the $2\rightarrow2$ processes only: 
in order to take into account this fact, at the same time avoiding the difficult implementation
of the radiative processes in the collision integral, we follow a very simple strategy, namely
we multiply the differential cross sections in 
Eqs.~\eqref{eq:1} and~\eqref{eq:2} by a temperature dependent overall factor
that allows us to reproduce the Arnold-Moore-Yaffe (AMY) production rate \cite{arnold2001} at a given temperature.
Within this implementation, we are sure that whenever the cell of the fluid is in local equilibrium at a temperature $T$,
the photon spectrum produced by that cell in our code is in fair agreement with the one implemented in calculations
based on hydro. Therefore what we implement in the collision integral are the following cross sections:
\begin{eqnarray}
\frac{d\sigma^{Compton}}{dt}&=&-\Phi(T)\frac{\pi\alpha\alpha_s}{3s^2}\frac{u^2+s^2}{us},
\label{eq:1a}
\\
\frac{d\sigma^{annihil}}{dt}&=&\Phi(T)\frac{8\pi\alpha\alpha_s}{9s^2}\frac{u^2+t^2}{ut},\label{eq:2a}
\end{eqnarray}
where $\Phi(T)$ is fixed by hand at the effective local temperature, 
$T = E/3N$, in order to reproduce the AMY rate, see below;
here $E$ is the energy of a given cell and
$N$ denotes the total number of particles in that cell.
The effective local temperature 
coincides with the actual temperature when the system is in local equilibrium;
on the other hand, when the system is out of equilibrium, 
we assume that the cross sections  are still given by Eqs.~\eqref{eq:1a} and~\eqref{eq:2a},
and $\Phi$ is computed at the scale  $T = E/3N$. We notice that
for large values of the effective temperature $\Phi\approx 1$ which means that for such large values
of the energy density, of the order of those expected in the initial stage, the cross sections implemented in the
collision integral are unaffected by this function, and the latter can be considered just as an tool to
reproduce the AMY rate when the system is in the equilibrated QGP phase.
The multiplicative function $\Phi$ does not depend on momenta: as we will show below, this simple choice is enough
to obtain a photon production rate that is in fair agreement with AMY in a quite broad range of temperature
and photon momentum. A plot of $\Phi(T)$ is shown in Fig.~\ref{Fig:phi}.

At this point it is important to clarify that, differently from the previous studies based on hydro,
we do not need to integrate the production rate over the spacetime volume of the system in order
to obtain the photon spectrum. As a matter of fact,
what we do is to implement photon production in the collision integral, by means of the  
microscopic cross sections in Eqs.~\eqref{eq:1a} and~\eqref{eq:2a}. In this way, we can follow
photon production consistently since the very  first moments after the collision, namely as soon as
the classical color fields decay and produce quarks and gluons, regardless of the fact that the system
is in local equilibrium or not.

\begin{figure}[t!]
\begin{center}
\includegraphics[scale=0.35]{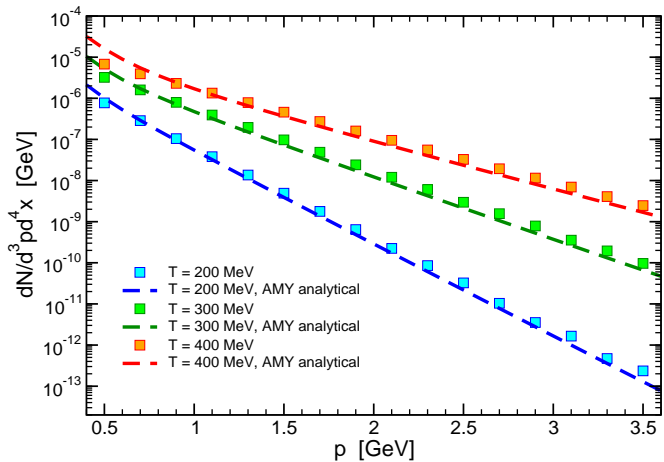}
\caption{\label{Fig:b1}Photon production rate for several values of the temperature.
Dashed lines correspond to the AMY rate at a given temperature, while squares represent
the rate implemented in our collision integral.}
\end{center}
\end{figure}

In Fig.~\ref{Fig:b1} we plot the photon production rate that we implement in the collision integral,
for three different temperatures: squares correspond to our rate, obtained by 
introducing a temperature dependent multiplicative factor in the $2\rightarrow2$ process rates 
in  order to increase the photon production rate,  while dashed data stand for the AMY rates
computed at the same temperature. We find that our procedure, although rough, reproduces the AMY rates fairly
well in the temperature range that is relevant for the RHICs.

\section{Results}

\subsection{Set up of the initial condition and QGP evolution}
In this subsection we discuss how we set up the initial classical field. 
We assume a longitudinal direction boost invariant  chromo-electric field at the initial time, 
and with a smooth profile in the transverse plane that mimics a standard Glauber-type distribution, namely
\begin{equation}
E_z^0(\bm x_T)=E^0_{\mathrm{max}}\left[\alpha~\rho_{\mathrm {coll}}(\bm x_T)
+(1-\alpha)\,\rho_{\mathrm{part}}(\bm x_T)\right],
\label{eq:45}
\end{equation}  
with $\alpha\leq1$, $\bm x_T$ denotes the transverse plane coordinate
and $\rho_{\mathrm {coll}},\rho_{\mathrm {part}}$ correspond to the
density of binary collisions and participants in the transverse plane respectively,
both being normalized to one.
In Eq.~\eqref{eq:45} two free parameters appear, namely the peak value of the magnitude
of the electric field $E^0_{\mathrm{max}}$  and the relative abundance of binary and participant
collisions $\alpha$.
These parameters are fixed in order to match a standard Glauber initialization: in particular,
for collisions at RHIC energy we impose that at $t= 0.6\,\mathrm{fm}/c$,
corresponding to a standard initialization time for calculations with the Glauber model,
the eccentricity of the system within the AFTm is equal to the eccentricity obtained within the Glauber model
with the same impact parameter, and we require that the total number of particles
produced by the two inizializations are the same; for collisions at LHC energy we 
perform the same tuning, by requiring that the eccentricity and the total multiplicity in the two initializations
coincide at $t= 0.3\,\mathrm{fm}/c$.
Numerical values of the parameters used in the calculations are in Table~\ref{Tab:1}.
We decide to implement these constraints on our initial condition because in this way
the main differences that we find in the photon spectrum  
can be related directly to the presence of the pre-equilibrium evolution
in the AFTm that instead is absent in the calculations with the Glauber initialization.

\begin{table}[t!]
\begin{tabular}{|c|c|c|c|}\hline
{\bf Collision}&{\bf Centrality}&$E_0$ [GeV$^2$]&$\alpha$\\
\hline
Au-Au, $200 A$GeV&$20-40$&$3.0$&$0.7$\\
\hline
Pb-Pb, $2.76 A$TeV&$20-40$&$6.0$&$0.85$\\
\hline
\end{tabular}
\caption{
\label{Tab:1}Parameters of Eq.~\eqref{eq:45} corresponding to the collisions examined in this study.}
\end{table}

\begin{figure*}[t!]
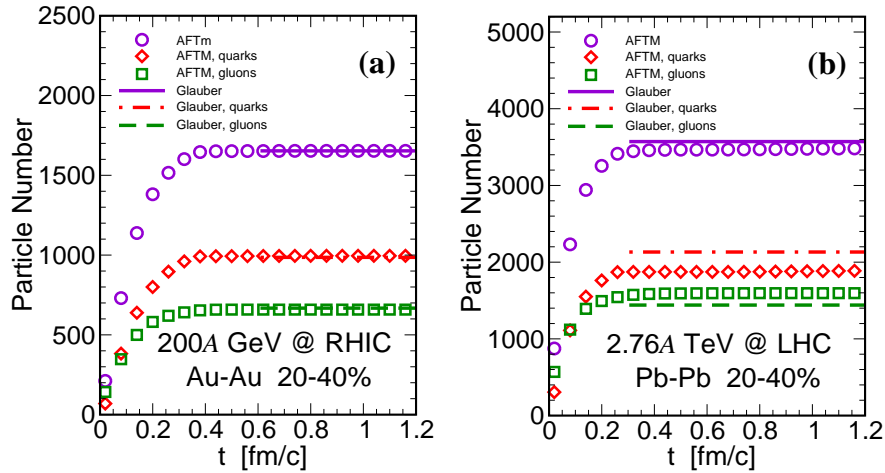

\begin{center}
\includegraphics[scale=0.35]{RHIC_bulk_mult_time.eps}~~
\includegraphics[scale=0.35]{LHC_bulk_mult_time.eps}
\caption{\label{Fig:c}Quark, gluon and total particle numbers for the AFTm initialization, compared with the
values used in the Glauber initialization. Panel {\bf (a)} corresponds to a collision at the RHIC energy
while panel {\bf (b)} stands for a collision at the LHC energy. 
Squares, diamonds and circles correspond to gluons, quarks and total particle number of the AFTm.
Both panels refer to collisions
belonging to the $20\%-40\%$ centrality class. }
\end{center}
\end{figure*}

In Fig.~\ref{Fig:c} we plot the particle numbers of quarks and gluons as a function of time, for the case of the
AFTm, and we compare these numbers with the ones we use in the calculations with the Glauber initialization.
Both panels refer to collisions in the $20\%-40\%$
centrality class, for collisions at RHIC in panel (a) and at LHC in panel (b).
In the Glauber case the multiplicity is chosen by matching it with the experimental
value of the $dN/dy$ for the given centrality class.
In both panels of Fig.~\ref{Fig:c}, the thick solid indigo lines denote the total number of particles we use in 
the simulation with the Glauber initialization, that should be compared to the
indigo  circles corresponding to the total number
of particles obtained within the AFTm initialization;  green thick dashed lines stand for the gluon number in Glauber
while the green squares correspond to gluon number in the AFTm. Finally, the thick red dot-dashed lines
correspond to quark+antiquark number in Glauber, while we use the red diamonds to
denote the same quantity for the AFTm. For the Glauber calculations we fix the ratio of quark+antiquark over
gluon numbers by its value at chemical equilibrium, that for massless particles is independent on temperature and depends
only on the degrees of freedom. 

The total particle numbers in AFTm and Glauber are the same by construction, 
while the relative abundance of quarks over gluons within the AFTm is not fixed a priori but it is a result of the
dynamical evolution of the system from the classical gluon field to the QGP via the Schwinger effect.
Nevertheless, we find that for the collision at RHIC not only the total particle number, but also the numbers of
quarks and gluons match those used in the Glauber model,
which are the chemical equilibrated ones. 
For the case of collisions at LHC we find some mismatch between the two initializations, even if the net difference is not very large.
We also notice that within the AFTm, quarks are produced very quickly for collisions at both
energies. In fact, starting from the classical color field that represents the gluon dominated initial state,
within 0.4 fm/c quarks are formed, and the relative abundance of quarks with respect to gluons is not very far from the 
one expected at chemical equilibrium, the latter being represented by the thick lines in Fig.~\ref{Fig:c}. 
%{\color{blue}
This is a bit different with what has been found in~\cite{Greif:2016jeb,Vovchenko:2016mtf,Ruggieri:2015tsa}, 
where although a gluon dominated state is considered in the initial condition, quarks are produced solely by
inelastic scatterings. This difference is clearly due to the fact that within our approach,
quarks and gluons are produced statistically on the same footing by the decay of the initial classical
gluon field. This difference affects photon production in the pre-equilibrium stage, as we discuss
in the next subsection.
%}

\begin{figure*}[t!]
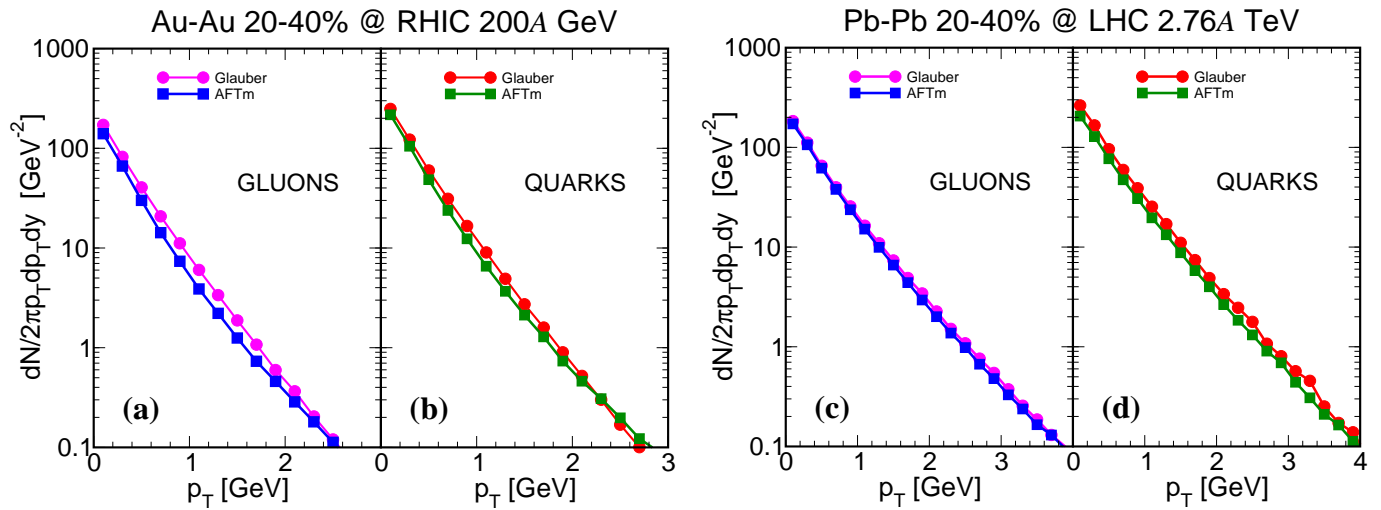

\begin{center}
\includegraphics[scale=0.35]{RHIC_bulk_spectra.eps}~~
\includegraphics[scale=0.35]{LHC_bulk_spectra.eps}
\caption{\label{Fig:d}Final spectra for gluons and quarks at RHIC (panels {\bf (a)} and {\bf (b)} respectively)
and at LHC (panels {\bf (c)} and {\bf (d)} respectively). In the panels,
squares correspond to AFTm while circles to Glauber calculations.}
\end{center}
\end{figure*}

In Fig.~\ref{Fig:d} we plot the final spectra for the QGP for the case of collisions at RHIC
(panels {\bf (a)} and {\bf (b)} respectively) and LHC (panels {\bf (c)} and {\bf (d)} respectively).
In the panels, squares correspond to AFTm while circles to Glauber calculations.
For both Glauber and AFTm initializations we add also the standard minijets, respectively
for $p_T \geq 2$ GeV in the case of RHIC collisions
and $p_T\geq 3$ GeV for LHC collisions.
We notice that 
final spectra of quarks and gluons in the Glauber calculations fairly agree with the ones
of the AFTm, both in the case of RHIC and LHC collisions.

\subsection{Sign of the early stage on photon spectrum and abundancy}

\begin{figure}[t!]
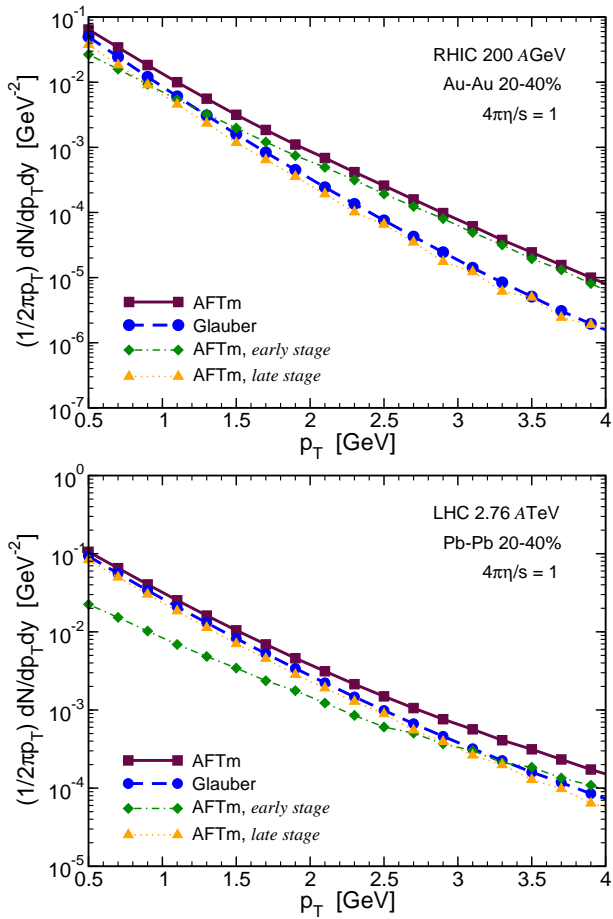

\begin{center}
\includegraphics[scale=0.3]{RHIC_photon_matching.eps}\\
\includegraphics[scale=0.3]{LHC_photon_matching.eps}
\caption{\label{Fig:e}Photon spectra at midrapidity. Panel {\bf (a)} corresponds to collisions at RHIC
and panel {\bf (b)} is the analogous for the LHC case.
In both panels, solid maroon lines correspond to the spectrum obtained within the AFTm, while blue dashed
lines stand for the spectrum obtained in simulations with the Glauber initialization.
Green dot-dashed lines denote the photon spectrum in the AFTm at $t_0=0.6$ fm/c for RHIC,
and $t_0=0.3$ fm/c for LHC: in both cases we call this the early stage spectrum. Finally, 
the orange dotted lines correspond to the difference between the maroon and the green lines,
that we call the late stage spectrum. See the text for more details.}
\end{center}
\end{figure}

In Fig.~\ref{Fig:e} we plot the photon spectra at midrapidity
for the $20\%-40\%$ centrality class. Panel {\bf (a)} corresponds to collisions at RHIC
and panel {\bf (b)} is the analogous for the LHC case.
In both panels, solid maroon lines correspond to the spectrum obtained within the AFTm, while blue dashed
lines stand for the spectrum obtained in simulations with the Glauber initialization.
Green dot-dashed lines denote the photon spectrum in the AFTm at $t_0=0.6$ fm/c for RHIC,
and $t_0=0.3$ fm/c for LHC: in both cases we call this the early stage spectrum. Finally, 
the orange dotted lines correspond to the difference between the maroon and the green lines,
that we call the late stage spectrum. 

Figure~\ref{Fig:e} corresponds to the main result of our study. Firstly we focus on the RHIC panel
since the results for LHC are in qualitative agreement with those for RHIC.
We start by noticing that the total number of photons in the case of the AFTm is larger than the
one obtained within the Glauber model. This is easy to understand: as a matter of fact,
photons in the AFTm are produced as soon as quarks and gluons appear by the decay
of the initial classical color field, while in the Glauber calculation this production is delayed up to 
the initialization time, the latter being usually assumed as the time necessary for the system to reach
a local equilibrium in the transverse plane. Integrating the photon spectrum over transverse
momentum and rapidity, we find that for the collisions at RHIC the photon abundancy in the AFTm
is approximately $30\%$ higher than that obtained within the Glauber model.
This difference, coming from the existence of a dynamics in the pre-equilibrium stage in the AFTm,
shows that pre-equilibrium photons are important as they give a substantial contribution to the total
number of photons produced by the QGP. Stated in other terms, the early stage is quite bright.
We have also computed the average temperature of the photon gas in the early stage
by means of the inverse slope of the photon spectrum: 
because of the pre-equilibrium dynamics, this quantity remains finite and in agreement with
the temperature of the bulk already computed in \cite{Ruggieri:2015yea}, 
instead of being divergent as it would be if it evolved as $T\propto\tau^{-1/3}$, 
that is as in the case of a thermalized system in a one-dimensional expansion.

Introducing a dynamics in the very early stage not only affects the total number of photons produced,
but also the shape of the spectrum.
In fact, in Fig.~\ref{Fig:e} we represent by the dotted green line the spectrum 
obtained at $t=0.6$ fm/c corresponding to the initialization time of the Glauber calculation;
we call this the early stage contribution to the photon spectrum.
We also plot the difference between the final AFTm spectrum and the early stage ones,
and represent this difference by the orange dotted line in Fig.~\ref{Fig:e}: we call this 
the late stage spectrum, corresponding to the photons produced by the QGP
since the initialization time of the Glauber simulation.
We notice that the late stage spectrum agrees with the one obtained
within the Glauber calculation: although the two initial conditions are different, 
the bulk evolution in the two models perfectly match each other starting from the
Glauber initialization time.
We would be tempted to name the late stage spectrum as the equilibrium  
spectrum: however, a strict distinction between equilibrium and pre-equilibrium
in the AFTm does not apply, because local equilibration in the fireball of the AFTm
takes place at different times in different cells, while in the Glauber model 
the spectrum is already equilibrated in the whole transverse plane at the initialization time.

\begin{figure}[t!]
\begin{center}
\includegraphics[scale=0.3]{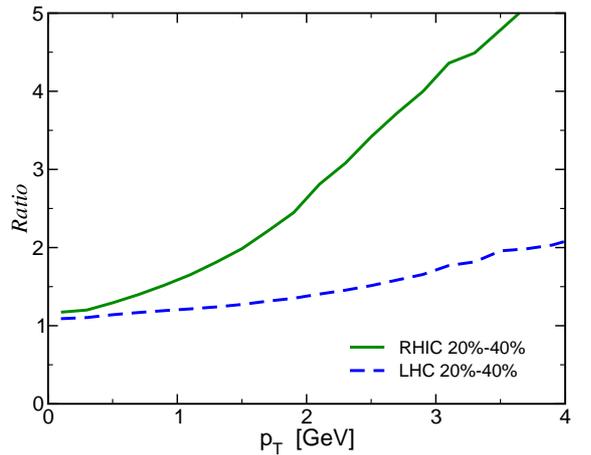}
\caption{\label{Fig:f}Ratio of photon spectrum of the AFTm over the one corresponding
to the Glauber calculation. Green solid line corresponds to RHIC collision and blue dashed line
to LHC collision.}
\end{center}
\end{figure} 
 
Comparing the green dot-dashed and the maroon solid lines in Fig.~\ref{Fig:e}, 
we notice that the most important photon production in the early stage
takes place in the momentum region
$p_T \gtrsim 1.5$ GeV. 
We identify the enhancement of the photon spectrum from QGP
in this momentum region as the sign of the early stage photons.

For a matter of comparison, 
we plot in Fig.~\ref{Fig:f} the ratio of the final photon spectrum of the AFTm over
the one corresponding to the Glauber calculation: the green solid line corresponds to 
collisions at RHIC while the blue dashed line to LHC collisions.
For what concerns Au-Au collisions, we notice the
enhancement of photon production within the AFTm for 
$p_T \gtrsim 1.5$ GeV with respect to the Glauber calculation.

The results for the LHC case, that are summarized in panel {\bf (b)} of Figure~\ref{Fig:e},
are qualitatively similar to those already discussed for the RHIC case. For collisions
at LHC we find that the domain $p_T \gtrsim 2$ GeV can be identified with the one in which
photons are produced in the very early stage. In this case we find that the photon abundancy
obtained within the AFTm is about the $20\%$ larger than the one obtained by the Glauber initialization,
thus the effect of the early stage seems to be smaller than the one observed in the case of RHIC collisions.
This can be easily understood because for collisions at LHC the early stage is considerably shorter than the
one at RHIC, and also the lifetime of the thermalized QGP at LHC is larger than the one at RHIC,
therefore the net effect of the early stage on photons produced by LHC collisions
is naturally smaller than the one that we have measured for RHIC collisions.
Therefore, we can summarize the results by stating that the early stage
affects considerably both photon abundancy and the shape of the photon spectrum,
because of the enhancement of photon production in the intermediate momentum region, 
namely $p_T \gtrsim 1.5$ GeV at RHIC and $p_T \gtrsim 2$ GeV at LHC.

\subsection{Comparison with PHSD and BAMPS}
\begin{figure}[t!]
\begin{center}
\includegraphics[scale=0.3]{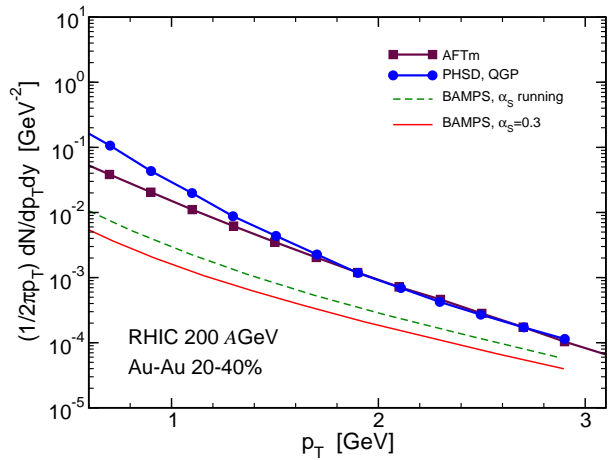}
\end{center}
\caption{\label{Fig:phsd}Photon spectrum produced by the quark-gluon plasma
for a RHIC collision,
within three different calculations based on relativistic transport: 
blue circles correspond to PHSD \cite{Linnyk:2015tha},
thin red solid line denotes the BAMPS result \cite{Greif:2016jeb} with a fixed strong coupling
while the dashed green line corresponds to the BAMPS result with a running couplingl;
finally the maroon squares correspond to our result.}
\end{figure}

%{\color{blue}
It is useful to compare our results for the photon spectrum with those obtained by means of other
calculations based on relativistic transport. This is done in Fig.~\ref{Fig:phsd} 
where we show the spectrum
of photons produced by the quark-gluon plasma within PHSD \cite{Linnyk:2015tha} (blue circles)
and BAMPS \cite{Greif:2016jeb} (red thin solid line, corresponding to a fix coupling,
and green dashed line corresponding to a running coupling). In the figure,
maroon squares correspond to our result.
%}

%{\color{blue}
The most striking aspect of the results shown in Fig.~\ref{Fig:phsd}  is that there is a disagreement
between PHSD and our results on the one hand, and BAMPS on the other hand. However, the reason of this discrepancy is very easy to understand: as a matter of fact, 
the BAMPS calculation uses a gluon dominated initial state but the conversion to quarks is
achieved only by the inelastic QCD scatterings, which has the effect to delay 
the appearance of quarks hence of the emission rate of photons. 
On the other hand, within the other two transport calculations quarks and gluons
are produced since the beginning because of field decay or string breaking, 
therefore leading to a larger photon emission of the quark-gluon plasma.

In last analysis, the results in Fig.~\ref{Fig:phsd} remind that nowadays there is still a theoretical uncertainty on the production time of the quark-gluon plasma in heavy ion collisions: one of the consequences of this uncertainty is measurable as the change of the photon abundancy of the quark-gluon plasma.
%}

\begin{figure}[t!]
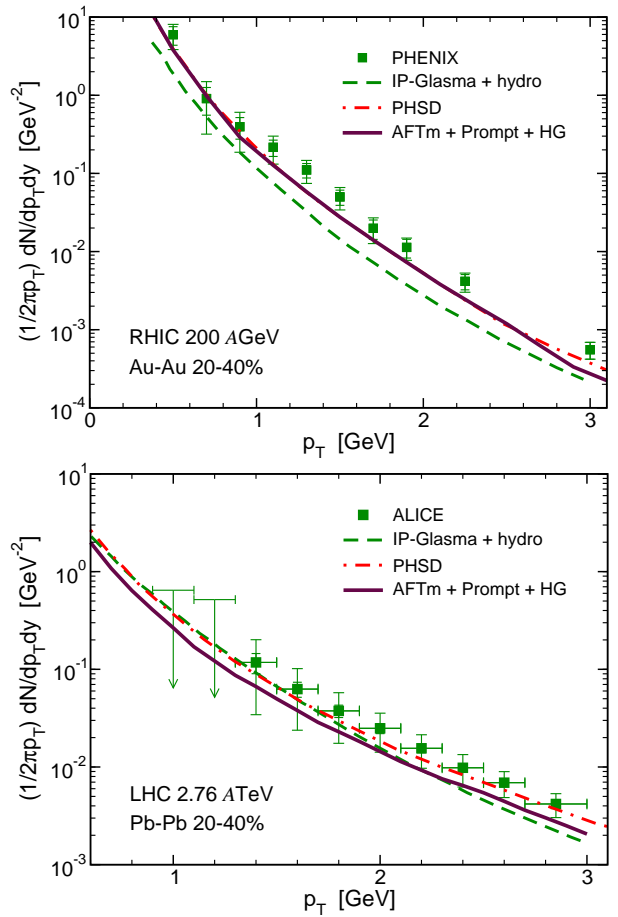

\begin{center}
\includegraphics[scale=0.3]{RHIC.eps}\\
\includegraphics[scale=0.3]{LHC.eps}
\end{center}
\caption{\label{Fig:rhic_all}In the upper panel we plot the direct photon spectrum for a RHIC collision.
Maroon solid line corresponds to our result for the quark-gluon plasma to which
we have added the prompt photons of~\cite{paquet2016} and the hadron gas
contribution of \cite{Linnyk:2015tha}.
For comparison, we have shown the IP-Glasma + hydro result
of~\cite{paquet2016}, represented by the green dashed line.
The red dot-dashed line denotes the PHSD result \cite{Linnyk:2015tha}.
Green squares correspond to the experimental data from the PHENIX collaboration \cite{Adare:2014fwh}.
In the lower panel we plot the direct photon spectrum for an LHC collision.
Line styles and colors are the same as in the upper panel. Experimental data
are from the ALICE collaboration \cite{Adam:2015lda}.}
\end{figure}

%{\color{blue}
A direct comparison with experimental data is not very fruitful at this stage of the work,
because in our code we miss the hadron gas contribution to photon spectrum.
Moreover, the difference between our results and those of other collaborations might be also come in part from the different energy density profiles and not only to the presence/absence of pre-equilibrium photons. A more detailed analysis is needed in order to compare quantitatively the theoretical predictions from the several collaborations. 
However, just to show that what we find for the quark-gluon plasma spectrum might lead the direct photon spectrum into the right ballpark, we add to our result
the prompt and the hadronic photons of  \cite{Linnyk:2015tha}.

In the upper panel of Fig.~\ref{Fig:rhic_all} we show the result for the direct photon spectrum for a RHIC collision.
The maroon solid line corresponds to our result obtained adding the prompt and hadronic photons
as explained above.
For comparison, we have shown the IP-Glasma + hydro result
of~\cite{paquet2016}, represented by the green dashed line.
The red dot-dashed line denotes the PHSD result \cite{Linnyk:2015tha}.
Experimental data are from the PHENIX collaboration \cite{Adare:2014fwh}.
We notice that the two transport calculations agree with each other remarkably
well, and that they tend to lower the tension with experimental data
with respect to the IP-Glasma + hydro calculation.
However, we do not push this result too much because
the hadronic contribution should be computed consistently rather than simply
added by hand: we plan to consider this problem in the future, while here we prefer
to focus on another problem namely the photon production from the pre-equilibrium stage. 

In the lower panel of Fig.~\ref{Fig:rhic_all} we show the result for the direct photon spectrum for an LHC collision.
We have once again added to our quark-gluon plasma photons the prompt and hadronic photons
computed in \cite{Linnyk:2015tha}. Experimental data are from the ALICE collaboration \cite{Adam:2015lda}.
In this case we find some minor disagreement between the three calculations,
which might be a consequence of the fact that for LHC collisions
the pre-equilibrium stage does not affect the direct photon spectrum in a relevant way. 

\section{Conclusions}
We have studied the photon production from quark-gluon plasma (QGP)
in relativistic heavy ion collisions (RHICs), putting emphasis on the role of the 
early stage quark-gluon scatterings on the direct photon spectrum.
We have considered a model for the initial conditions in RHICs,
based on a classical gluon field mimicking the Glasma, beside a mechanism for the conversion
of the field to QGP; the dynamics of the QGP 
has been studied by means of a simulation code based  on relativistic transport theory
coupled to the classical field dynamics. We have simulated
both Au-Au collisions at RHIC energy and Pb-Pb collisions at LHC energy.
Our approach, although simplified with respect to the more realistic situation based
on Glasma and its evolution,
allows us to follow consistently the system since the very initial stage up to the freezout.

Within our theoretical description,
QGP is produced since the very early stage by the decay of the classical gluon field by means of the
Schwinger mechanism. As soon as QGP is formed, 
quarks and gluons scatter and they can produce photons: our approach therefore allows us
to compute the contribution of the QGP to the photon spectrum taking into account
also the ones produced in the early stages, that have been neglected in previous studies.
Although in the collision integral we have considered only the $2\rightarrow 2$ 
photon production processes, we have artificially modified the scattering matrix of these 
processes in order to reproduce the celebrated AMY rate, see Fig.~\ref{Fig:b1}:
as a consequence, 
our production rate agrees with the one commonly used in calculations based on hydro.

We have been able to identify a transverse momentum region 
in which direct photon spectrum is dominated by the early stage photons,
namely $p_T \gtrsim 1.5$ GeV for collisions at RHIC and 
$p_T \gtrsim 2$ GeV for collisions at LHC, see Figg.~\ref{Fig:e} and~\ref{Fig:f}.
Moreover, we have found that during the early stages the amount of photons
produced is approximately within the $20-30\%$ of the total amount of photons produced
by the QGP. This is a remarkable result considering that
the lifetime of the early stage is at most one tenth of the full QGP lifetime in the fireball:
we can conclude that the early stage is quite bright, or stated in other terms, 
that there is no dark age in RHICs.

%{\color{blue}
We have tried a tentative comparison of our results with the existing RHIC and LHC data about 
the direct photon spectrum, borrowing the hadronic and prompt photon contributions
from existing works \cite{Linnyk:2015tha}. We have found that the net result is in fair agreement
with the existing data, as well as with other relativistic transport calculations \cite{Linnyk:2015tha,Greif:2016jeb}. 
We have found some disagreement with BAMPS \cite{Greif:2016jeb}, which however is clearly understood
as arising from both a different initial condition and a different initial early stage dynamics. 
%}

For a matter of simplicity we have not included here the 
Bose-Einstein enhancement factors in the collision integral, which are potentially relevant
in the early stage due to the large gluon occupation numbers; these have been considered in 
a recent study \cite{Berges:2017eom} and we plan to include them in future works 
following \cite{Scardina:2014gxa}.
%{\color{blue}
It would also be important to consider initial state fluctuations that have been proved to
be important for photon production~\cite{Chatterjee:2013naa,Chatterjee:2012dn},
as well as to gauge the collision integral to the holographic production rate \cite{Iatrakis:2016ugz}
rather than to the AMY rate. 
%In addition to these problems, it will be important to consider a nonabelian dynamics \cite{Voronyuk:2015ita} instead of the abelian flux tube that we have used here.
These important upgrades of our calculations will be the subject of forthcoming works.
We have only considered the direct photon spectrum in this work; due to the importance of the photon elliptic flow, in relation also to the solution of the direct photon puzzle, we will devote a detailed analysis of this quantity and its comparison to experimental data in future studies.
%}

\begin{acknowledgements}
V. G., S. P. and F. S. would like to acknowledge the ERC-STG funding under a QGPDyn grant. 
M.R., G.X.P and V.G would like to thank the National Natural Science Foundation of China (11575190 and 11475110) and the CAS Present’s International Fellowship Initiative (2015PM008 and 2016VMA063). 
We thank Jean-Fran\c{c}ois Paquet, Raju Venugopalan, Zhe Xu and Moritz Greif for discussions and correspondence.
\end{acknowledgements}

\end{document}